\newcommand{\ba}{BaCu$_2$Si$_2$O$_7$}
\newcommand{\kcuf}{KCuF$_3$}

\documentclass[aps,prl,twocolumn,superscriptaddress]{revtex4}
\usepackage{graphicx}

\begin{document}

\title{Dominance of excitation continuum in the longitudinal spectrum of weakly coupled Heisenberg $S=1/2$
chains.}
\author{A. Zheludev}
\email[]{zhelud@bigfoot.com}

\affiliation{Solid State Division, Oak Ridge national Laboratory,
Oak Ridge, TN  37831-6393, USA.}

\author{K. Kakurai}
\affiliation{Advanced Science Research Center, Japan Atomic Energy
Research Institute, Tokai, Ibaraki 319-1195, Japan.}

\author{T. Masuda}
\affiliation{\label{ut}Department of Advanced Materials Science,
The University of Tokyo, Tokyo 113-8656, Japan.}

\author{K. Uchinokura}
\affiliation{\label{ut}Department of Advanced Materials Science,
The University of Tokyo, Tokyo 113-8656, Japan.}

\author{K. Nakajima}
\affiliation{Institute of Solid State Physics, The University of
Tokyo, Tokai, Ibaraki 319-1106, Japan.}

\date{\today}

\begin{abstract}
A low-field spin flop transition in the quasi one-dimensional
antiferromagnet \ba\ is exploited to study the polarization
dependence of low-energy magnetic excitations. The measured
longitudinal spectrum is best described as single broad continuum,
with no sharp ``longitudinal mode'', in apparent contradiction
with the commonly used chain-MF/RPA theories. The observed
behavior is also quite different than that previously seen in the
related \kcuf\ material, presumably due to a large difference in
the relative strength of inter-chain interactions. The results
highlight the limitations of the chain-MF/RPA approach.
\end{abstract}

\maketitle \narrowtext

The physics of weakly-interacting spin chains is of great
fundamental interest, as it represents the crossover from quantum
to semi-classical spin dynamics. Excitations in conventional
three-dimensional magnets are single-particle spin wave states.
These can be described as precessions of magnetic moments around
their equilibrium orientation, and are necessarily polarized
perpendicular to the order parameter. In contrast, the excitation
spectrum of one-dimensioanl (1D) quantum-disordered
antiferromagnets (AFs) is isotropic.  In particular, in the
$S=1/2$ Heisenberg model, the spectrum is a
polarization-independent multi-spinon continuum
\cite{Fadeev81,Haldane93,KCUF3,Tennant95-II,Dender96}. Arbitrary
weak inter-chain interactions in {\it quasi-1D} $S=1/2$ AFs
restore long-range order (LRO)
\cite{Scalapino75,AffleckGelfand94,AffleckHalperin96,Wang97},
break rotational invariance and reinstate the transverse spin
waves at low frequencies \cite{Schulz96,Essler97}. The continuum
persists in the high-frequency regime.

 A counter-intuitive novel feature of
weakly-interacting Heisenberg $S=1/2$ chains is the {\it
longitudinal mode} (LM), a long-lived ``spin wave'' polarized {\it
parallel} to the direction of ordered moment
\cite{Schulz96,Essler97}. This new type of excitation was recently
discovered in the \kcuf\ compound \cite{Lake00,Nagler}, where it
is seen as a single symmetric broad peak in the longitudinal
spectrum.\cite{Lake00} Such behavior is not fully understood
within the commonly used chain-Mean-Field \cite{Scalapino75} and
Random Phase Approximation (MF/RPA) theories
\cite{Schulz96,Essler97} that predict a sharp (long-lived) LM and
a separate broad longitudinal continuum (LC). To further probe the
limitations of the MF/RPA approach, and learn more about the
exotic LM, we performed a polarization-sensitive inelastic neutron
scattering study of \ba. The advantage of this material is a
considerably smaller ratio of inter-chain and in-chain
interactions, as compared to \kcuf. This brings us closer to the
idealized weak-coupling limit, where the MF/RPA theory can be
expected to work best. Surprisingly, the measured longitudinal
spectrum substantially deviates from theoretical predictions.
Contrary to expectation, the LM in \ba\ is altogether ill-defined,
and the low-energy longitudinal spectrum is best described as a
single broad continuum feature. Thus, for longitudinal
polarization, there is no analogue of the separation of
single-particle and continuum states, that was previously seen in
the transverse spectrum \cite{ZheludevKenzelmann00,
ZheludevKenzelmann01}.

The magnetism of \ba\ is by now very well characterized using bulk
methods \cite{Tsukada99,Tsukada01}, neutron diffraction
\cite{Tsukada99,ZheludevRessouche02} and inelastic neutron
scattering
\cite{Tsukada99,ZheludevKenzelmann00,KenzelmannZheludev01,ZheludevKenzelmann01}.
The material is orthorhombic (space group $Pnma$, $a=6.862$~\AA,
$b=13.178$~\AA, $c=6.897$~\AA ) with slightly zigzag AF $S=1/2$
chains of Cu$^{2+}$ ions running along the $c$ axis. The in-chain
exchange constant is $J=24.1$~meV. Interactions between the chains
are much weaker, and the characteristic badwidth of spin wave
dispersion perpendicular to the chain direction (``mass gap'' in
the terminology of Ref.~\cite{Essler97}) is $\Delta=2.51$~meV.
\ba\ orders antiferromagnetically at $T_{N}=9.2$~K$=0.033
J/k_\mathrm{B}$ with a zero-$T$ saturation moment of only
$m_{0}=0.15$~$\mu_\mathrm{B}$. For comparison, for \kcuf,
$J=37$~meV, $T_{N}=39$~K$=0.09J/k_\mathrm{B}$,
$m_{0}=0.54$~$\mu_\mathrm{B}$
(Ref.~\cite{Satija80,KCUF3,Tennant95-II} and references therein).
Due to a weak easy-axis aniso1tropy in \ba, the ordered staggered
magnetization is aligned along the $c$ axis.

The most direct way to determine the polarization of magnetic
excitations is by using polarized neutrons. Such measurements are
typically associated with severe intensity losses in the
polarizing monochromator and analyzer. Given the size of currently
available single-crystal samples, and the expected scattering
cross section for longitudinal excitations in \ba, this type of
experiment does not appear feasible. Fortunately though, even for
an unpolarized neutron beam, the scattering intensity is
polarization dependent. Only spin components perpendicular to
momentum transfer contribute to the cross section. One commonly
used strategy is to measure the inelastic signal at equivalent
wave vectors in different Brillouin zones (BZs). Comparing such
data sets can provide polarization information, but relies heavily
on the exact knowledge of the spectrometer resolution function.
The latter can be estimated quite accurately, but an error of only
a few percent can ruin the polarization analysis. This was the
main problem in previous inconclusive attempt to apply this
technique to longitudinal excitations in \ba, as described in
Ref.~\cite{ZheludevKenzelmann01}.

\begin{figure}
\includegraphics[width=3.0in]{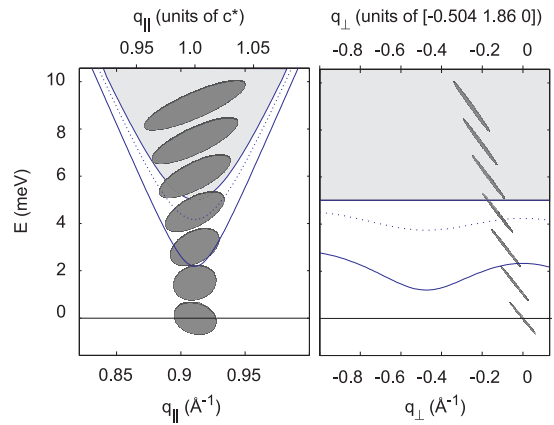}
\caption{$(E-\mathbf{q})$-space trajectory of the inelastic scan
and the corresponding evolution of the FWHM resolution ellipsoid
(shaded ellipses). The solid curve shows the dispersion of spin
waves in \ba. The dotted line and light shaded area represent the
dispersion of the longitudinal mode and the domain of the
excitation continuum in \ba, respectively, as predicted by the
MF/RPA model.\label{trajec}}
\end{figure}

In the present work the problem was overcome by using the
``reverse'' approach. The {\it same} inelastic scan (always in the
same BZ) was performed for {\it different} orientations of the
ordered moment in the system. Polarization information was then
extracted in a point-by-point data analysis that is robust and
insensitive to any resolution effects. As described in detail
below, to change the direction of ordered moment we exploited one
of the recently discovered field-induced spin-re orientation
transitions in \ba\ \cite{ZheludevRessouche02}. A 5~mm diameter
and 40~mm long cylindrical single crystal sample was mounted on
the the thermal 3-axis spectrometer PONTA (5G) at the JRR3-M
reactor at Japan Atomic Energy Research Institute at Tokai. The
sample cylinder axis is close to the $(1,1,0)$ reciprocal-space
direction and was oriented vertically in the experiment. The
$(0,0,1)$ and $(-0.5,1.86,0)$ reciprocal-lattice vectors
determined the horizontal scattering plane. Sample environment was
a horizontal-field 6~T cryomagnet. The field was at all times
applied along the $c$ axis of the crystal. The data were collected
in an energy scan close to the 1D AF zone-center $(0,0,1)$. The
design of the horizontal-field magnet imposes stringent
constraints on experimental geometry, allowing only narrow $\pm
15^{\circ}$ windows for the incident and scattered beams. For this
reason, each data set was collected while simultaneously scanning
the energy transfer $\hbar \omega$ and momentum transfer
perpendicular to the chain axis $\mathbf{q}_\bot$, at a constant
momentum transfer along the chains $q_\|=2\pi/c$, which is the 1D
AF zone-center. Figure~\ref{trajec} shows the trajectory of the
scan relative to the previously determined spin wave dispersion in
\ba. Also shown is the evolution of the instrument resolution
ellipsoid, calculated in the Cooper-Nathans approxomation
\cite{Cooper67,Popovici75}.

\begin{figure}
\includegraphics[width=3.2in]{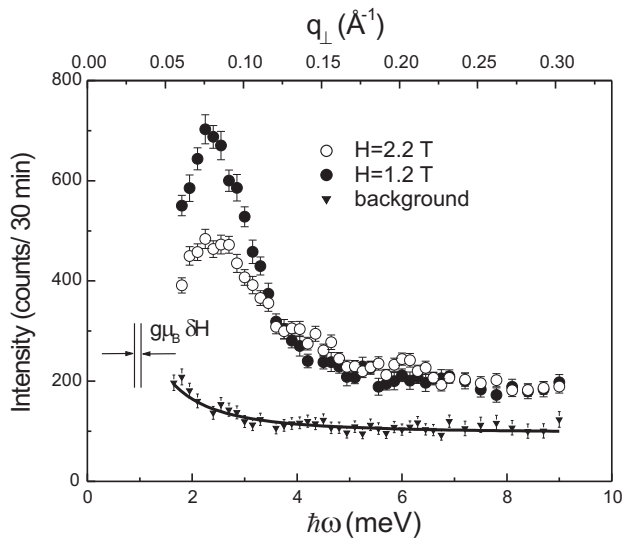}
\caption{Inelastic scans measured in \ba\ near the 1D AF
zone-center $(0,0,1)$ at $T=2$~K in magnetic fields applied along
the $c$-axis. The data were collected below (solid circles) and
above (open circles) the spin-flop transition at $H_c=2.0$~T. The
background (triangles) was measured as described in the text. The
solid line is a Lorentzian fit to the background
scan.\label{data}}
\end{figure}

\begin{figure}
\includegraphics[width=3.2in]{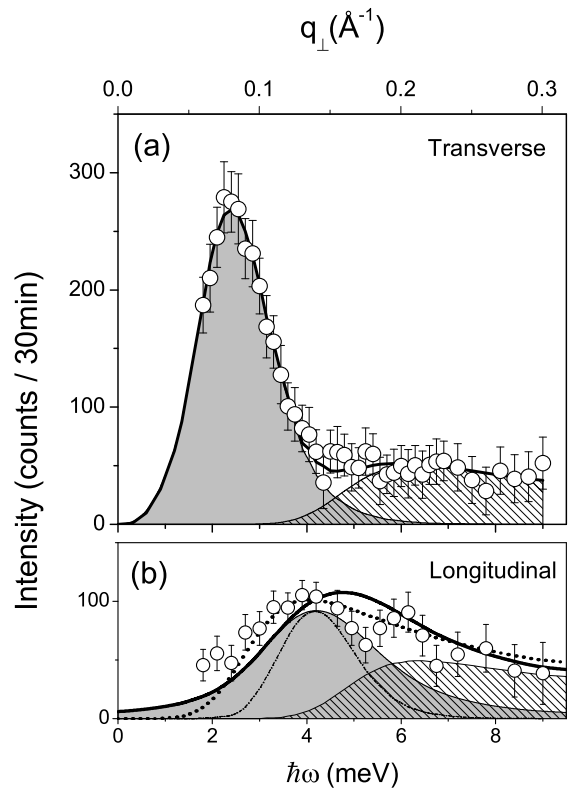}
\caption{ Transverse- (a) and longitudinal-polarized (b) dynamic
structure factor measured in \ba near the 1D AF zone-center. The
heavy solid line is a fit based on MF/RPA theory, as described in
the text. The grayed and hatched areas represent the single-mode
and continuum parts of the cross section. In (b) the heavy dotted
line is a fit that does not include any single-mode contribution,
as described in the text. The thin dash-dot line is the
experimental energy resolution.\label{result}}
\end{figure}

The scan was repeated  in three independent measurements. The
first measurement (Fig.~\ref{data}, solid circles) was performed
in a magnetic field $H_1=1.2$~T at $T=2$~K. Under these
conditions, just as at $H=0$, the ordered moment in \ba\ is
parallel to $c$ \cite{KenzelmannZheludev01,ZheludevRessouche02}.
The measured intensity is related to the longitudinal (parallel to
the ordered moment) and transverse (perpendicular to the ordered
moment) dynamic structure factors $S^{\|}(\mathbf{q},\omega)$ and
$S^{\bot}(\mathbf{q},\omega)$ through
 $I_{1}(\mathbf{q},\omega)\propto
 S^{\bot}(\mathbf{q},\omega)(1+\cos^2\alpha)+S^{\|}(\mathbf{q},\omega)\sin^2\alpha
 +\mathcal{B}(\mathbf{q},\omega).$
 Here $\alpha$ is the angle between the momentum transfer
$\mathbf{q}$ and the chain axis, and
$\mathcal{B}(\mathbf{q},\omega)$ is the background, assumed to be
of non-magnetic origin. In our particular case $\alpha\lesssim
15^\circ$ through the entire scan, and it is mostly the transverse
spin fluctuations that are observed. In the a second measurement
the same scan was repeated at $H_2=2.2$~T, which exceeds the
critical field $H_{c}=2.0$~T of a spin-flop transition
\cite{Tsukada01,ZheludevRessouche02}. These data are plotted with
open symbols in Fig.~\ref{data}. In the spin-flop state the
ordered staggered moment is perpendicular to the field direction
and therefore lies within the crystallographic $(a,b)$ plane. The
measured inelastic intensity can be written as $
I_{2}(\mathbf{q},\omega)\propto
 S^{\bot}(\mathbf{q},\omega)(1+\sin^2\alpha)+S^{\|}(\mathbf{q},\omega)\cos^2\alpha
 +\mathcal{B}(\mathbf{q},\omega).\label{two}$
 For small $\alpha$ one observes an equal mixture of longitudinal
and transverse spin fluctuations. In the final measurement,
performed at zero field, the sample was removed from the magnet
and sample holder. The same scan was repeated to directly measure
the extrinsic background $\mathcal{B}$, mainly due to spurious
scattering in the sample environment (triangles in
Fig.~\ref{data}). The intrinsic background (from scattering in the
sample itself) was previously shown to be negligible
\cite{ZheludevKenzelmann01}. At each $(\mathbf{q},\omega)$-point
the equations for $I_1$ and $I_2$ were then solved to extract the
dynamic structure factors for different polarizations, as plotted
in Fig.~\ref{result}.

Using this method to determine the polarization of spin
fluctuations is based on several assumptions. The first is that
the spin structure rotates {\it as a whole} when going through the
spin-flop transition, but is otherwise unaffected. For \ba, such
behavior was confirmed by previous neutron diffraction studies in
magnetic fields up to 5~T \cite{ZheludevRessouche02}. In
particular, upon crossing $H_c$, the magnitude of the ordered
moment  remains almost constant, and the change in relative
alignment of nearby spins is insignificant. The second requirement
is that magnetic anisotropy and Zeeman effects are negligible on
the experimentally relevant energy scales, as defined by the gap
$\Delta$ and the experimental energy resolution (typically 1.7~meV
FWHM). In our measurements this condition is also satisfied.
Indeed, at $H=H_2=2.2$~T the Zeeman energy per spin is an order of
magnitude smaller: $gS\mu_\mathrm{B}H\sim 0.25$~meV. Performing
differential measurements at $H_1$ and $H_2$, with
$gS\mu_\mathrm{B}\delta H$ of only 0.13~meV,  further minimizes
the effect, as shown by arrows in Fig.~\ref{data}. The magnitude
of magnetic anisotropy in \ba\ is also negligible, the associated
spin wave gap being only about 0.2~meV
\cite{ZheludevKenzelmann01}. In summary, the spin flop transition
only changes the coordinate system in which the terms
``longitudinal'' and ``transverse'' are defined, but should not
affect the nature of the excitation spectrum.

Previous measurements of {\it transverse} spin fluctuations in
\ba\ \cite{ZheludevKenzelmann00,ZheludevKenzelmann01} were found
to be in remarkably good quantitative agreement with MF/RPA theory
\cite{Schulz96,Essler97}. In this approach each individual spin
chain is viewed as subject to an effective staggered exchange
field $\mathbf{H}_\pi$, generated by LRO neighboring chains. The
staggered field in turn generates an attractive potential between
spinons, producing 2-spinon bound states. Two components of the
spin-triplet bound state are polarized perpendicular to
$\mathbf{H}_\pi$. These are analogues of conventional spin waves,
and have a gap $\Delta \propto H_\pi^{2/3}$.  The third component
is the longitudinal mode with a gap $\Delta_{\|}=\sqrt{3}\Delta$.
Regardless of polarization, a multi-magnon continuum, the remains
of the spinon-continuum, has a gap of $\Delta_c=2\Delta$. Starting
from this MF solution the RPA is used to calculate the spin wave
dispersion perpendicular to the chains. For \ba\ the MF/RPA
dynamic structure factor for transverse spin waves is explicitly
written out in Eqs.~4--7 in Ref.~\cite{ZheludevKenzelmann01}. The
continuum part is well approximated by the ``truncated Muller
ansatz'' functional form, as given by Eqs.~10 and 11 in the same
reference. As an important consistency check, this model cross
section function, after a numerical convolution with the
spectrometer resolution, was fit to the transverse spectrum
measured in this work. Since all parameters of the model were
previously determined with very good accuracy
\cite{KenzelmannZheludev01,ZheludevKenzelmann01}, the only
parameter varied in the present data analysis was an overall
intensity scaling factor. An excellent fit was obtained, and is
shown in a solid line in Fig.~\ref{result}a. Continuum and single
mode parts are shown a hatched and greyed areas, respectively.

We now  turn to discussing the main result of this work, namely
the measured longitudinal spectrum. As shown in
Fig.~\ref{result}b, the scattering is a broad asymmetric peak with
a maximum around 4~meV energy transfer, and an extended ``tail''
on the high-energy side. As a first step, we analyzed the data
using a model cross section based on MF/RPA theory, as for
transverse excitations. The corresponding dynamic structure factor
for the LM is given by Eqs.~12 and 13 in
Ref.~\cite{ZheludevKenzelmann01}. At the particular wave vector,
the MF/RPA predicts the LM at 4.2~meV, quite close to the observed
intensity maximum. Given the shape of the observed peak, in our
fits the LM was allowed to have a non-zero intrinsic energy width.
The corresponding $\delta$-function in Eq.~12 of
Ref.~\cite{ZheludevKenzelmann01} was replaced with a Lorenzian
profile of FWHM $\Gamma$. In addition, the prefactor $\gamma$ that
chracterizes the energy-scaled intensity of the longitudinal mode
relative to that of transverse spin waves was treated as an
adjustable parameter. Continuum scattering was assumed to be
exactly as for transverse polarization. This model, when
convoluted with the spectrometer resolution function, yields a
reasonably good two-parameter fit to the data with $\gamma=2.2(1)$
and $\Gamma=2.4(7)$~meV (heavy solid line in Fig.~\ref{result}b).

The obtained parameter values clearly indicate the failure of the
MF/RPA model to describe longitudinal excitations in \ba. First,
the refined intensity coefficient $\gamma$ is about 4.5 times
larger, than the MF/RPA prediction \cite{Essler97}. Second, to
obtain a good fit, we had to assume a substantial intrinsic width
for the LM. In contrast, the LM in the MF/RPA model is infinitely
sharp. For {\it weakly}-coupled chains like those in \ba, this
damping effect has a particularly important consequence: the LM
becomes altogether ill-defined. Indeed, in this case, the
continuum is relatively strong, and comparable in intensity to
single-mode scattering. In the MF/RPA the LC has a (pseudo)gap at
$2\Delta$, and the LM is centered at about $1.73 \Delta$. At the
same time, our analysis shows that in \ba\ the intrinsic FWHM of
the LM is itself be about $\Delta$. Given these numbers, the LM
and the LC {\it are too close to be separated}, and constitute
{\it a single continuum feature}. To illustrate this fact, in
Fig.~\ref{result}b we also show  a fit to the longitudinal data
using {\it only} the truncated Muller ansatz (heavy dotted line),
with no single mode contribution. With a continuum gap
$\Delta_c=1.4(1)\Delta$ (smaller than the expected MF/RPA value of
$2\Delta$), this empirical function describes the data at least as
well as the cross section that also includes a damped LM. The
situation is quite different for materials with inter-chain
interactions of intermediate strength, such as \kcuf. Here the LC,
which is the high-energy ``tail'' seen in constant-$q$ scans, is
relatively weak. As a result, even though the LM is broadened, it
is seen as a well-defined symmetric inelastic peak\cite{Lake00}.

Comparing the spectra measured in \ba\ and \kcuf, we conclude that
the MF/RPA picture based on distinct LM and LC contributions does
{\it not} become more accurate in the weak-coupling limit. The
width of the LM peak normalized by $\Delta$ in the weaker-coupled
\ba\ system is even greater, than in the \kcuf\ compound. This
counter-intuitive behavior may be related to the critical nature
of spin correlations in isolated $S=1/2$ chains. Quantum
criticality implies that there is no characteristic energy scale
in the chains, and therefore {\it any} inter-chain coupling is to
be considered as {\it strong}. Strictly speaking, the RPA is well
justified only for inter-chain interactions small compared to the
energy gap in individual chains. In our case, the chains are
intrinsically gapless. The gap $\Delta$ only emerges in the MF
theory and is necessarily of the same order of magnitude as
inter-chain coupling itself \cite{Schulz96}. The problem is
therefore only marginally suited for the RPA for intrinsic
reasons.

In summary, despite a small ratio of inter-chain and in-chain
interactions, the low-energy longitudinal spectrum of \ba\  is
dominated by a broad excitation continuum, and there appears to be
no distinct ``longitudinal mode''. Further theoretical
calculations beyond the MF/RPA are needed to better understand
this behavior.

Work at the University of Tokyo was supported in part by the
Grant-in-Aid for COE Research ``SCP coupled system" of the
Japanese Ministry of Education, Culture, Sports, Science, and
Technology. Oak Ridge National Laboratory is managed by
UT-Battelle, LLC for the U.S. Department of Energy under contract
DE-AC05-00OR22725. We would like to thank F. Essler, S. Maslov and
A. Tsvelik (BNL) for iluminating discussions.


\end{document}